\documentstyle[prd,floats,psfig,aps]{revtex}
\begin{document}
\title{Collision of spinning black holes in the close limit}
\author{Gaurav Khanna}
\address{Natural Science Division,\\
Southampton College of Long Island University,\\
Southampton NY 11968.}

\maketitle
\begin{abstract}
In this paper we consider the collision of spinning holes using first order 
perturbation theory of black holes (Teukolsky formalism). With these results 
(along with ones, published in the past) one can predict the properties of the 
gravitational waves radiated from the late stage inspiral of two spinning, 
equal mass black holes. Also we note that the energy radiated by the 
head-on 
collision of two spinning holes with spins (that are equal and opposite) 
aligned along the common axis is more than the case in which the spins are 
perpendicular to the axis of the collision.
\end{abstract}

\section{Introduction}

There is considerable current interest in studying the collision of
two black holes, since these events could be primary sources of
gravitational waves for interferometric gravitational wave detectors
currently under construction. The mathematics describing such events  
consists of the nonlinear partial differential equations of general 
relativity, Einstein's theory of gravitation. These equations are very 
difficult to solve especially in the case of two merging holes, or of 
the highly distorted final hole that is formed by the merger. Limited 
(in resolution and evolution time) full numerical simulations of grazing 
collisons of black holes are being done now \cite{potsdam}, but long term 
stable evolutions 
are still distant in the future. It is therefore of interest to have at 
hand approximate results which in certain regimes could be used to test 
and possibly even complement the full numerical evolutions. Among such 
approximation 
methods is the ``close-limit" approximation in which spacetime of a black 
hole 
collision is represented as a single distorted black hole and evolutions are 
done with simple linear perturbative equations. In the past, this method 
has been applied to the case of a head-on collision of two boosted holes 
\cite{boost} and the case of slow inspiral \cite{njp} with 
considerable success. 
In both these cases, the black holes considered had no spin. In this work, we 
will consider the collision of spinning holes. 

Since we are attempting a linear, first order perturbative calculation, it is 
sufficient to study the coalescence of two spinning holes with zero initial 
linear momentum. This is because, to obtain the results for the
inspiral of spinning holes, one can simply do a linear superposition of
these results with those from our past work \cite{njp}. However, the first 
order perturbative treatment imposes several restrictions 
on the physical scenarios we can treat. In addition to the close-slow limitations, 
we can only easily analyze cases in which the spins of the two black holes are 
equal and opposite. This happens because, to lowest order, the (perturbative) 
extrinsic curvature of a black hole with spin $S$ and a conformal distance $L$ 
from the origin is of order $SL$. Superposing another hole with parallel and 
equal spin and at $-L$ will yield a situation with zero perturbation in 
extrinsic curvature, which is not a case of interest.

So in the following, we shall present two cases that are exhaustive. 
{\it Case I}, when the two
spins are equal and opposite and aligned along the common axis of the two 
holes. Specifically, we  place the
two holes at $x=\pm L/2$ and have their spins parallel and anti-parallel to the
x-axis. {\it Case II}, when the two spins are equal and opposite and
perpendicular to the common axis of the two holes. Here, we place the two holes 
at $x=\pm L/2$ and have their spins parallel and anti-parallel to the y-axis. 
These cases have no net angular momentum. So we expect Zerilli and Teukolsky 
formalisms to agree exactly (Kerr parameter $a=0$). We, therefore report 
calculations and results from the Teukolsky formalism only.
Part of this work, in the context of the Zerilli formalism can be found in
this reference \cite{screw}.

\section{Initial data}

To evolve a spacetime in general relativity, one needs to provide
initial data, a 3-geometry $g_{ab}$ and an extrinsic curvature
$K_{ab}$, that solve Einstein's equations on some starting
hypersurface (i.e., at some starting time). These initial value equations 
have the form,
\begin{eqnarray}
\nabla^a (K_{ab} - g_{ab} K) &=& 0\\
{}^3R-K_{ab} K^{ab} + K^2 &=&0
\end{eqnarray}
where $g_{ab}$ is the spatial metric, $K_{ab}$ is the extrinsic
curvature and ${}^3R$ is the scalar curvature of the three metric. If
we propose a 3-metric that is conformally flat $g_{ab} = \phi^4
\widehat{g}_{ab}$, with $\widehat{g}_{ab}$ a flat metric, and $\phi^4$
the conformal factor, and we use a decomposition of the extrinsic
curvature $K_{ab} = \phi^{-2} \widehat{K}_{ab}$, and assume maximal
slicing $K_a^a=0$, the constraints become,
\begin{eqnarray}
\widehat{\nabla}^a \widehat{K}_{ab} &=& 0\label{momentum}\\
\widehat{\nabla}^2 \phi &=& -\frac{1}{8}
\phi^{-7} \widehat{K}_{ab} \widehat{K}^{ab}\ ,\label{hami}
\end{eqnarray}
where $\widehat{\nabla}$ is a flat-space covariant derivative.

To solve the momentum constraint, we start with a solution that 
represents a single hole with spin $S$ \cite{CoYo},
\begin{equation}
\hat{K}^{\rm one}_{ab} = {3 \over r^3} \left[{\epsilon_{cad}} {S^d}{n^c}{n_b} 
+{\epsilon_{cbd}}{S^d}{n^c}{n_a}\right]\ .
\end{equation}
In this expression for the conformally related extrinsic curvature at
some point $x^a$, the quantity $n_b$ is a unit vector, in the ``base''
flat space with metric $\hat{g}_{ab}$, directed from a point
representing the location of the hole to the point $x^a$. The symbol
$r$ represents the distance, in the flat base space, from the point
of the hole to $x^a$. It is straightforward to show that 
the solution of the Hamiltonian constraint corresponding to eq. (5)
corresponds to a spacetime with  ADM angular momentum $S^{a}$.

The next step is to modify this to represent holes centered at 
$x=\pm L/2$ in the conformally flat metric. Since the 
momentum constraint is linear, we can simply add two expressions
of the above form,
\begin{equation}
\widehat{K}^{\rm two}_{ab} =
\widehat{K}^{\rm one}_{ab}\left(x \rightarrow x-L/2,{S_{x}\, {\rm or} \, S_{y}} = S \right)+
\widehat{K}^{\rm one}_{ab}\left(x \rightarrow x+
L/2,{S_{x}\, {\rm or} \, S_{y}} =-S \right)\ .
\end{equation}
We will choose
in further expressions to use a polar coordinate system in the flat
space determined by $\hat{g}_{ab}$
centered in the mid-point separating the two holes and label the polar
coordinates as $(R,\theta,\phi)$. So $R$ will be the distance in the
flat space from the midpoint between the holes.

To solve the Hamiltonian constraint \ref{hami}, we introduce an
approximation, (the slow approximation) which we will show is enough
for our purposes. In fact, in this approximation the solution for the
conformal factor turns out to be the familiar Misner \cite{Mi}
solution if one chooses the topology of the slice to have a single
asymptotically flat region, or the Brill--Lindquist \cite{BrLi}
solution if there are three asymptotically flat regions.

\subsection{The slow approximation}

We assume that the black holes are initially close, and that the
initial spins $S$ are small. We denote by $\vec{n}^+$ and
$\vec{n}^-$ the normal vectors corresponding, respectively, to the one
hole solutions at $x=+L/2$ and at $x=-L/2$, and we recall $R$ to be
the distance to a field point, in the flat conformal space, from the
point midway between the holes.  For large $R$, the normal vectors
$\vec{n}^+$ and $\vec{n}^-$ almost cancel.  More specifically
$\vec{n}^+=-\vec{n}^-+O(L/R)$. A consequence of this is that the total
initial $\widehat{K}^{ab}$ is first order in $L/R$. It scales linearly with 
the spin $S$ as well. Thus the source term in the Hamiltonian constraint 
is quadratic in $S$. If we choose to find a solution to the conformal factor to
first order in $S$ (which should give us a good approximation in the
case of slowly spinning holes), we can ignore this quadratic source
term. So now, the Hamiltonian constraint looks like the one for zero
momentum, which is simply the Laplace equation. A well known solution
to this, is the Misner solution \cite{Mi}. This solution, is
characterized by a parameter $\mu_0$ which describes the separation of
the two throats.  We can relate this parameter to the conformal
distance $L$ in the following way \cite{NCSA},
\begin{equation}
L/M=\frac{\rm{\coth}\mu_0}{2\Sigma_1}\ \ \hspace*{30pt}
  \ \ \ \Sigma_1\equiv\sum_{n=1}\frac{1}{\sinh n\mu_0}\ .
\end{equation}

We must now map the coordinates of the initial value solution to the
coordinates for the Schwarzschild (Kerr with $a=0$) background. To do this, 
we interpret the $R$  as the isotropic radial coordinate of a Schwarzschild 
spacetime, and we relate it to the usual Schwarzschild radial coordinate $r$ by
$R=(\sqrt{r}+\sqrt{r-2M})^2/4$.  From this we arrive at the following
expressions for the components of the metric and extrinsic curvature.

\subsection{Initial data for the Teukolsky function -- Case I}

Following the construction as outlined in the last section,
we get the following expressions for the initial spatial metric and 
extrinsic curvature:
\begin{equation}
{K}_{ab} = {3 S L \over R r^{2}} \left[
\begin{array}{ccc}0&\frac{-2\sin\theta\sin 2\varphi}{\sqrt{1-2M/r}}&
    \frac{-2\sin\theta\sin 2\theta\cos^{2}\varphi}{\sqrt{1-2M/r}}\\
\frac{-2\sin\theta\sin 2\varphi}{\sqrt{1-2M/r}}&r\cos\theta\sin 2\varphi&
r\sin\theta(\cos^{2}\theta\cos^{2}\varphi-\sin^{2}\varphi)\\
\frac{-2\sin\theta\sin 2\theta\cos^{2}\varphi}{\sqrt{1-2M/r}}&
r\sin\theta(\cos^{2}\theta\cos^{2}\varphi-\sin^{2}\varphi)&
-r\sin^{2}\theta\cos\theta\sin 2\varphi
\end{array}\right]\ .
\end{equation}
The metric has a form, identical to the Misner solution \cite{Mi}.
Now, we use the methodology and expressions in \cite{njp}, to find 
the initial data for the Teukolsky function, $\Psi={\rho}^{-4}\psi_4$,
where ${\rho}={-1/{(r-ia\cos\theta)}}$. 

For the azimuthal modes $m=\pm 2$ we get these expressions,
\begin{equation}
-{\Psi\over\sqrt{2\pi}}=-{3\over 4}i(\cos\theta\pm 1)^{2}LS\frac{(r-2M)
(-1+2\sqrt{1-{2M\over r}})}{r(r(1+\sqrt{1-{2M\over r}})-M)}
\end{equation}
\begin{equation}
-{\dot\Psi\over\sqrt{2\pi}}=-{3\over 4}i(\cos\theta\pm
1)^{2}LS\frac{(r-2M)
(-3r+8M+3(r-7M)\sqrt{1-{2M\over r}})}{r^{3}(r(1+\sqrt{1-{2M\over r}})-M)}.
\end{equation}

And for the azimuthal mode $m=0$ we get,
\begin{equation}
-{\Psi\over\sqrt{2\pi}}={3\over 2}i\sin^{2}\theta LS\frac{(r-2M)
(-1+2\sqrt{1-{2M\over r}})}{r(r(1+\sqrt{1-{2M\over r}})-M)}
\end{equation}
\begin{equation}
-{\dot\Psi\over\sqrt{2\pi}} ={3\over 2}i\sin^{2}\theta LS\frac{(r-2M)
(-3r+8M+3(r-7M)\sqrt{1-{2M\over r}})}{r^{3}(r(1+\sqrt{1-{2M\over r}})-M)}.
\end{equation}

To all these expressions, we also need to add metric contributions. They
have the same form as the ones in \cite{njp}, therefore we will not list 
them here. 

\subsection{Initial data for the Teukolsky function -- Case II}

Again, following the construction as outlined in the last 
section, and using the close-slow approximation, we arrive at the following 
expressions for the initial spatial metric and extrinsic curvature:
\begin{equation}
{K}_{ab} = {3 S L \over R r^{2}} \left[
\begin{array}{ccc}\frac{2\cos\theta}{r-2M}&\frac{\sin\theta
(4\cos^{2}\varphi-1)}{\sqrt{1-2M/r}}&\frac{-\sin 2\theta 
\sin\theta\sin 2\varphi}{\sqrt{1-2M/r}}\\
\frac{\sin\theta(4\cos^{2}\varphi-1)}{\sqrt{1-2M/r}}&
-2r\cos\theta\cos^{2}\varphi&{r\over2}\sin\theta(1+\cos^{2}\theta)
\sin 2\varphi\\
 \frac{-\sin 2\theta \sin\theta\sin 2\varphi}{\sqrt{1-2M/r}}&
{r\over2}\sin\theta(1+\cos^{2}\theta)\sin 2\varphi&
-{r\over2}\sin\theta\sin 2\theta\sin 2\varphi 
\end{array}\right]\ .
\end{equation}
The metric again has a form, identical to the Misner solution \cite{Mi}.
Using the methodology and expressions we discussed in \cite{njp}, 
the initial data for the Teukolsky function, is:

For the azimuthal modes $m=\pm 2$ we get,
\begin{equation}
-{\Psi\over\sqrt{2\pi}}=\mp{3\over 4}(\cos\theta\pm 1)^{2}LS\frac{(r-2M)
(-1+2\sqrt{1-{2M\over r}})}{r(r(1+\sqrt{1-{2M\over r}})-M)}
\end{equation}
\begin{equation}
-{\dot\Psi\over\sqrt{2\pi}}=\mp{3\over 4}(\cos\theta\pm
1)^{2}LS\frac{(r-2M)
(-3r+8M+3(r-7M)\sqrt{1-{2M\over r}})}{r^{3}(r(1+\sqrt{1-{2M\over r}})-M)}.
\end{equation}

And for the azimuthal mode, $m=0$ the initial data are identically zero. 
To all these expressions we need to add the metric contributions, that
have a form identical to the ones in our past work \cite{njp}. 

\section{Evolution of the Data using the Teukolsky equation}

Given the Cauchy data from the last section, the time evolution is
obtained from the Teukolsky equation \cite{Te},
\begin{eqnarray}
&&
\Biggr\{\left[a^2\sin^2\theta-\frac{(r^2 + a^2)^2}{\Delta}\right]
\partial_{tt}-
\frac{4 M a r}{\Delta}\partial_{t\varphi}
+ 4\left[r+ia\cos\theta-\frac{M(r^2-a^2)}{\Delta}\right]\partial_t
\nonumber\\
&&+\,\Delta^{2}\partial_r\left(\Delta^{-1}\partial_r\right)
+\frac{1}{\sin\theta}\partial_\theta\left(\sin\theta\partial_\theta\right)
+\left[\frac{1}{\sin^2\theta}-\frac{a^2}{\Delta}\right]
\partial_{\varphi\varphi}\\
&&-\, 4 \left[\frac{a (r-M)}{\Delta} + \frac{i \cos\theta}{\sin^2\theta}
\right] \partial_\varphi
-\left(4 \cot^2\theta +2 \right)\Biggr\}\Psi=0,
\end{eqnarray}
where $M$ is the mass of the black hole, $a$ its angular momentum per
unit mass (which is zero in our case), $\Sigma\equiv 
r^2+a^2\cos^2\theta$, and $\Delta\equiv
r^2-2Mr+a^2$. Evolving the initial data we just calculated, with this equation
will enable us to extract gravity wave waveforms that correspond to the late 
stage merger of spinning holes. We can also estimate the energy carried away by 
these gravitational waves. The radiated energy is given by \cite{CaLo},
\begin{equation}
\frac{dE}{dt}=\lim_{r\to\infty}\left\{ \frac{1}{4\pi r^{6}}
\int_{\Omega}d\Omega\left| \int_{-\infty}^{t}d\tilde{t}\ \Psi(\tilde{t
},r,\theta,\varphi) \right|^2\right\}, \quad
d\Omega=\sin\theta\ d\vartheta\ d\varphi.
\end{equation}
Note that the angular momentum radiated in the cases we consider is zero. 

\section{Results of the evolutions}

In this section we show waveforms and plots for energy radiated 
from the collision of two spinning holes. Recall that the two
holes have equal mass and equal and opposite spin, and we are
considering two cases, {\it Case I} in which the spins are aligned 
along the common axis of the two holes, and {\it Case II} in which
the two spins are perpendicular to the common axis of the two holes.

The waveforms that follow, are for a collision of two
black holes that were initially separated by a conformal distance of
$0.91$ and had an individual spin of $0.1$ in units of ADM
mass. 

\begin{figure}
\centerline{\psfig{figure=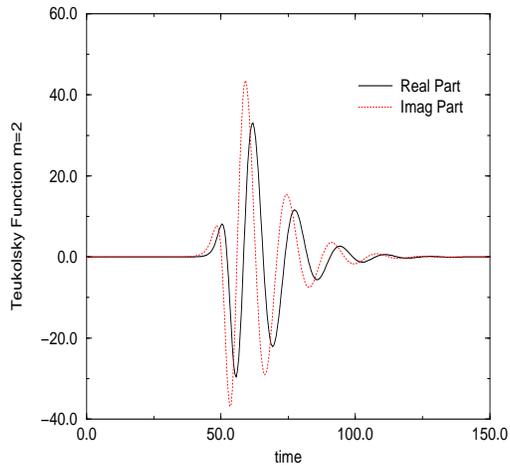,height=72mm,width=72mm}}
\caption{This figure depicts the $m=2$ mode of the Teukolsky function 
for a Case I collision. The solid line depicts the real part whereas the dotted
one, the imaginary part. All quantities are in units of ADM mass. 
}
\label{waveformI}
\end{figure}

\begin{figure}
\centerline{\psfig{figure=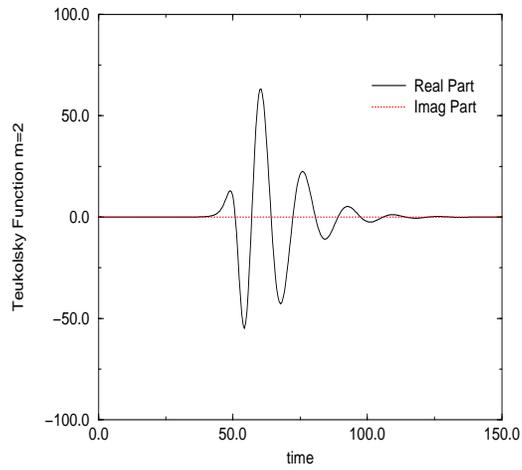,height=72mm,width=72mm}}
\caption{This figure shows the $m=2$ mode of the Teukolsky function 
for a Case II collision. The solid line depicts the real part whereas the 
dotted one, the imaginary part. All quantities are in units of ADM mass. }
\label{waveformII}
\end{figure}

In figure \ref{waveformI} we show the $m=2$ mode of the Teukolsky function
for a {\it Case I} collision. We see the typical quasi-normal ringing, in
both the real and imaginary parts of the function. In
figure \ref{waveformII} we plot the $m=2$ mode of the Teukolsky function
for a {\it Case II} collision. Note that the initial data is purely real
for this case, hence the imaginary part of the waveform is identically zero. 
Quasi-normal ringing is self-evident here as well. This is the type of 
signal that gravity wave observatories like LIGO, will detect if they 
happen to witness a collision of the kind we are considering. 

\begin{figure}
\centerline{\psfig{figure=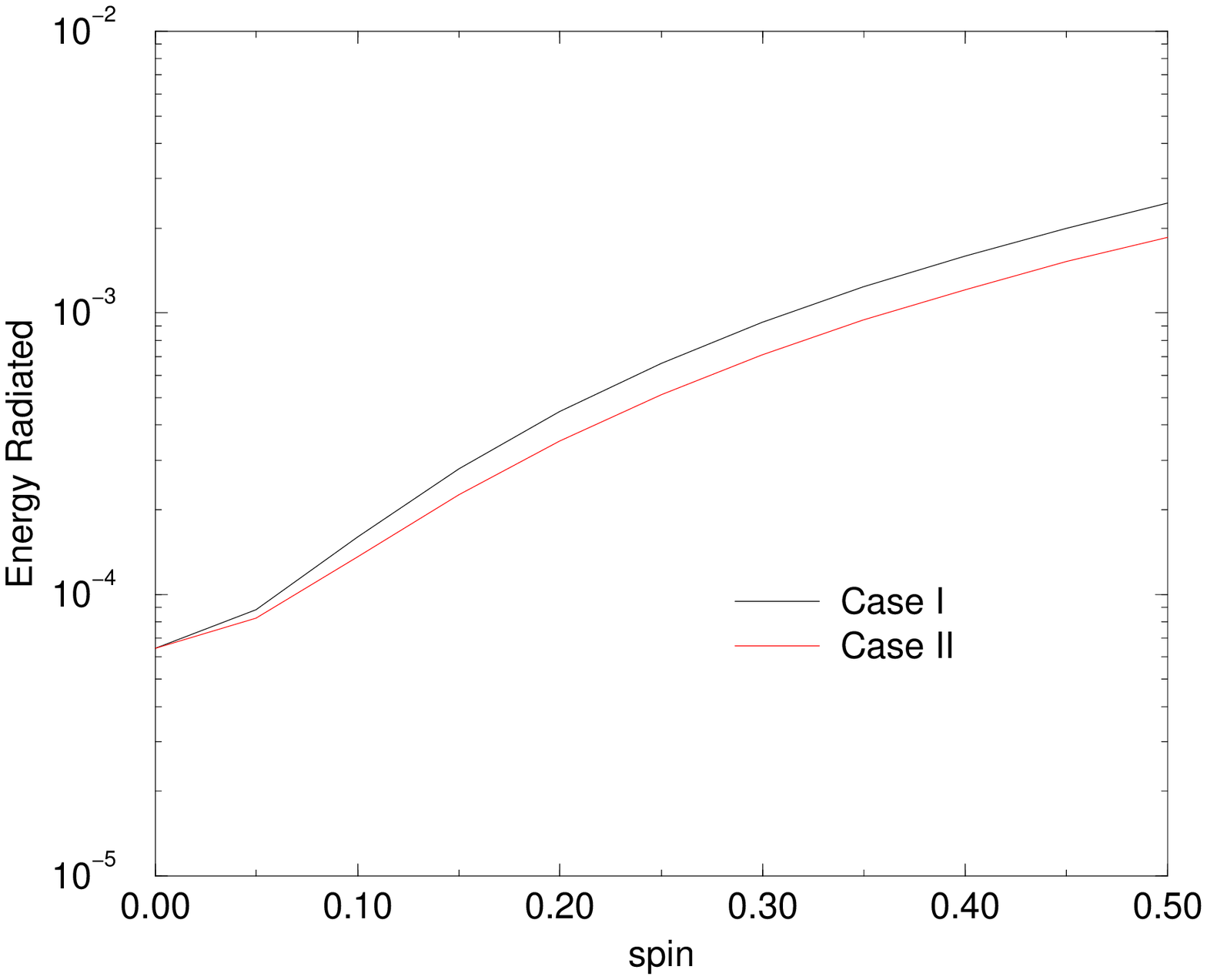,height=72mm,width=72mm}}
\caption{The radiated energy in a head-on collision of two
spinning black holes as a function of the individual initial spin
(that is equal and opposite for the two holes), for a fixed conformal 
separation of $0.91$. All quantities are in ADM mass units. 
Looking at the behavior of the curves for high spin, one can estimate 
that less than $1\%$ of the mass of the system is lost in a typical 
collision involving spinning holes.}
\label{energy}
\end{figure}

Let us turn now to the evaluation of the radiated energies for these
two cases. Figure \ref{energy} shows the radiated energy as a function of
the initial spin, for a fixed separation of the holes. The first 
noteworthy thing is that the two curves, corresponding to {\it Case I}
and {\it Case II} are very close to each other throughout. This 
indicates that the geometry of the initial configuration does not
matter much as far as energy loss is concerned. But, it should be observed 
that for any spin $S$, {\it Case I} radiates more than {\it Case II}. The next 
important observation (by looking at the trend of the curves for large spin) is
that, even for high values of the initial spin of the two holes this
type of collision radiates less than $1\%$ of its total mass. This
means that inclusion of spin does {\it not} dramatically change
our original estimate \cite{njp} \cite{prl} of a percent of the mass-energy 
being carried away by gravity waves in binary black hole inspiral. 

\section{Conclusions}

We performed a first order, perturbative calculation to study the merger of
two spinning holes. The exhaustive cases considered had holes with spins either 
aligned or perpendicular to the common axis of the two holes. For these cases
we extracted waveforms and estimated energy radiated. We observed that in the 
case in which the spins of the two holes are perpendicular to the common axis,
the system radiates a little less than the case in which the spins are along
the common axis. In either case, the energy radiated appears to be a fraction of
a percent of the total mass of the system. Since we know that the late stage
inspiral of two non-spinning holes typically radiates about a percent of the 
mass-energy \cite{prl} \cite{njp} this implies that adding spin to the
holes doesn't really change such estimates by much. 

With these results (along with our earlier ones \cite{njp}) we can predict 
the properties of the gravity waves radiated from the ringdown phase of the
slow inspiral of two spinning, equal mass black holes. 

\section{Acknowledgments}

The author is grateful for research support and equipment from Long Island 
University. Part of this work was done at the Pennsylvania State 
University with support from NSF grants PHY-9800973 and PHY-9423950. The 
author thanks J. Pullin, P. Laguna, R. Gleiser, D. Cartin, D. Arnold and 
L. Smolin for comments and suggestions.

\end{document}